\documentstyle[11pt]{article}
\newcommand{\Ca}{C_{\rm A}}
\newcommand{\Cf}{C_{\rm F}}
\newcommand{\Tf}{T_{\rm F}}
\newcommand{\Nl}{N_{\rm L}}
\newcommand{\ep}{\varepsilon}
\newcommand{\om}{\omega}
\newcommand{\qq}{\lv\bar q q\rv}
\newcommand{\as}{\alpha_{\rm s}}
\newcommand{\wh}{\widehat}
\newcommand{\wt}{\widetilde}
\newcommand{\Nc}{N_{\rm c}}
\newcommand{\oc}{\om_{\rm c}}
\newcommand{\lv}{\mbox{$<$}}
\newcommand{\rv}{\mbox{$>$}}
\begin{document}
\thispagestyle{empty}
\begin{center}
\vspace*{\fill}
{\Large\bf Operator product expansion in static-quark effective\\[5pt]
field theory: large perturbative correction}
\vfill
{\large D.~J.~BROADHURST$^*$}
\vskip.5cm
{\large Physics Department, \,Open University, \,Milton Keynes MK7 6AA, \,UK}
\vskip.7 cm
{\large A.~G.~GROZIN$^+$}
\vskip.5cm
{\large Institute of Nuclear Physics, \,630090 Novosibirsk, \,USSR}
\end{center}
\vfill
{\bf Abstract }We calculate the coefficients of operators with dimensions
$d\le7$ in the operator product expansion of correlators of $\bar q\Gamma
Q$ currents, for the effective field theory of an infinite-mass quark, $Q$.
Exact two-loop results are obtained, with an arbitrary gauge group and
spacetime dimension, for the perturbative ($d=0$) and quark-condensate
($d=3$) contributions, confirming our previous result for the anomalous
dimension of the current. Leading-order results are given for light-quark
operators with $d=5,6,7$ and gluon operators with $d=4,6$. The existence of
a perturbative correction of order 100\% precludes a reliable determination
of $f_{\rm B}$ from non-relativistic sum rules.
\vfill
\hfill September 30, 1991\\
{\footnotesize $^*$ d\_broadhurst@vax.acs.open.ac.uk}\\
{\footnotesize $^+$ grozin@vxinpc.inp.nsk.su}\hfill{OUT-4102-31}
\newpage
\setlength{\parindent}{0.7cm}


\section{Introduction}
\label{Intro}

The effective field theory (EFT) of a static quark~\cite{eah,eft}
considerably simplifies the QCD analysis of hadrons containing a single
heavy quark. In particular, mesons in EFT are analogous to the hydrogen
atom in QED, rather than to positronium. To leading order in $1/M$, EFT
possesses a heavy-quark spin symmetry~\cite{wis}, enabling hadrons to be
classified according to the angular momentum and parity, $j^\pi$, of light
fields. The $Q\bar q$ ground state with $j^\pi = \frac12^-$ ($S$\/-wave
antiquark) gives $0^-$ and $1^-$ mesons with identical properties ($m_{\rm
B}=m_{{\rm B}^*}$, $f_{\rm B}=f_{{\rm B}^*}$). Excited states with $j^\pi =
\frac12^+$ and $\frac32^+$ ($P$\/-wave antiquark) give degenerate pairs of
mesons $0^+$, $1^+$ and $1^+$, $2^+$.  The splitting in each pair (i.e.\
the hyperfine structure) is of order $1/M$.

In QCD, however, even such a simplified problem requires nonperturbative
methods, such as lattice simulation~\cite{lat}. An alternative method is
provided by sum rules~\cite{svz} based on the operator product expansion
(OPE).

In order to investigate meson properties in EFT, we consider correlators of
bilinear currents, $\wt J=\bar q\Gamma\wt Q$, where the tildes distinguish
EFT quantities from those of conventional QCD and the static-quark field
satisfies $\wt Q=\gamma_0\wt Q$.  The irrelevance of $\gamma_0$ reduces the
time-like component of the (pseudo)vector current to the (pseudo)scalar
current; $\sigma_{0i}$ to $\gamma_i$; $\sigma_{ij}$ to
$\epsilon_{ijk}\gamma_k\gamma_5$.  There are thus four such currents in
EFT, with $\Gamma=\gamma_5$, $\gamma_i$ and $\Gamma=1$, $\gamma_i\gamma_5$.
The quantum numbers of the first pair, with $\Gamma$ anticommuting with
$\gamma_0$, are those of the ground-state $j^\pi=\frac12^-$ mesons; the
quantum numbers of second pair, with $\Gamma$ commuting with $\gamma_0$,
are those of the excited-state  $j^\pi=\frac12^+$ mesons.  Currents with
quantum numbers of other mesons necessarily involve derivatives.

In the $\overline{\rm MS}$ scheme, the renormalized current $\wt J(\mu)$ is
obtained from the bare current $\wt J_0=\bar q_0\Gamma\wt Q_0$, in
$D\equiv4-2\ep$ dimensions, by the multiplicative renormalization $\wt
J(\mu)=\bar\mu^{2\ep}\wt Z_J^{-1}(\mu)\wt J_0$, with
$\bar\mu^2=\mu^2e^\gamma/4\pi$. The renormalization constant $\wt Z_J$ was
calculated at the one-loop level in~\cite{vas} and to two loops
in~\cite{ef1,jam}. The EFT currents $\wt J$ are related to the
corresponding QCD currents $J$ at the one-loop level by~\cite{eah}
\begin{eqnarray}
\wt J(\mu) &=& J(\mu) \Biggl[1 + \Cf\frac{\as(\mu)}{4\pi}
\left(\frac{H^2-10}{4}\log\frac{M^2}{\mu^2} - \frac34 H^2 + H H'
\mp \frac12 H + 4\right)
\nonumber\\
&+&O\left(\as^2(\mu)\right)\Biggr]+O\left(\frac1M\right),
\label{EFT/QCD}
\end{eqnarray}
where $\gamma_\mu\Gamma\gamma_\mu=H\Gamma$, $H'=\partial H/\partial D$, and
in this and all subsequent equations we adopt the convention that the upper
sign corresponds to  $\Gamma$ anticommuting with $\gamma_0$ and the lower
sign to the commuting case. Each EFT current, $\bar q\Gamma\wt Q$, is
related to two QCD currents, $\bar q\Gamma Q$ and $\bar q\Gamma\gamma_0 Q$,
by formula~(\ref{EFT/QCD}), which we have verified by methods simpler than
those in~\cite{eah}, using dimensional regularization of infrared
singularities, as in~\cite{mz2}.

In co-ordinate space, the correlator
$i\lv T\wt J(x)\wt J^\dagger(0)\rv$
is proportional to $\delta({\bf x})$.
Its Fourier transform
\begin{equation}
\wt\Pi(q_0) = i\int dx\;e^{iqx}
\lv T\wt J(x)\wt J^\dagger(0)\rv
\label{Corr}
\end{equation}
therefore depends only on the energy $q_0$ and is given by the dispersion
relation:
\[\wt\Pi(\om) = \int\limits_0^\infty
\frac{\wt\rho(\om')d\om'}{\om'-\om}         
-\cdots\,,\]
where the dots denote a quadratic subtraction polynomial. The contribution
of the lowest meson, M, is
$\wt\rho_{\rm M}(\om)=\wt f_{\rm M}^2 \delta(\om-\om_{\rm M})$, where
$|\lv0|\wt J(\mu)|{\rm M}\rv|=\wt f_{\rm M}(\mu)$ is its coupling, and $\om_{\rm
M}$ its binding energy.  Note that the EFT meson state $|{\rm M}\rv$
lacks the relativistic normalization factor $\sqrt{2E}$ of QCD
and that $\om$ and $\om_{\rm M}$, like all energies in EFT,
are measured relative to the pole mass $M$ of the heavy quark.

The correlator~(\ref{Corr}) has the structure
\begin{equation}
\wt\Pi(\om)=
{\rm Tr}                                    
\left[ \Gamma \frac{1+\gamma_0}2 \bar\Gamma 
\left\{\gamma_0A(\om)+B(\om)\right\}\right] 
=2\left\{A(\om)\mp B(\om)\right\},
\label{Struct}
\end{equation}
where $\bar\Gamma=\gamma_0\Gamma^\dagger\gamma_0$ and $A$ and $B$ derive
from operators with even and odd dimensions, respectively.
Thus the correlators of $\bar q\gamma_5\wt Q$ and $\bar q\gamma_i\wt Q$
coincide, yielding the same ground-state sum rule.  Similarly, the
correlators of $\bar q\wt Q$ and $\bar q\gamma_i\gamma_5\wt Q$ coincide,
yielding an excited-state sum rule obtained by a change of sign of the
odd-dimensional contributions.

Mesons with a single heavy quark have been considered in applications of
nonrelativistic~\cite{shu,czz}, relativistic
Borel-transform~\cite{aae,gen,may}, and moment-method~\cite{rry,bag,dap}
QCD sum rules. (A comparison is made in~\cite{nar}.) EFT sum rules are
similar to nonrelativistic~\cite{shu} sum rules, but in the QCD case the
perturbative corrections are plagued by powers of a large hybrid logarithm:
$\log(M/\om)$. In EFT these are summed, {\em ab initio}, by the
introduction of a new anomalous dimension, leading to well defined
radiative corrections, whose size has so far not received due attention.
Here we remedy that state of affairs.

In Sections~\ref{Pert} and~\ref{QCond} we calculate the perturbative
($d=0$) and quark-condensate ($d=3$) contributions, up to two loops, using
the method proposed in~\cite{ef1}. These contributions should dominate the
sum rule. In the Section~\ref{Higher} we give higher-dimensional
contributions, up to $d=7$, to leading order. A Borel-transform EFT sum
rule is derived in Section~\ref{Concl} and conclusions are drawn from a
numerical investigation.  Throughout, we omit $1/M$ corrections, though
these may be systematically included.  We believe that a proper account of
the radiative corrections in the $M\to\infty$ limit is required first, in
order to determine whether the ``QCD hydrogen atom'' is amenable to
sum-rule analysis. Our conclusion has importance for any more realistic
analysis of the B meson.


\section{Two-loop perturbative contribution}
\label{Pert}

In~\cite{ef1} we gave a method for calculating two-loop EFT diagrams, with
one external momentum and zero light-quark mass, and implemented it as a
REDUCE~\cite{red} package that evaluates these diagrams in terms of the
basic structures $\Gamma_{1,2}$, where
\[\Gamma_n\equiv\left[(-2\om)^{-2\ep}\Gamma(-\ep)/(4\pi)^{D/2}\right]^n
\Gamma(1+2n\ep).\]
Using this package, we evaluate the bare perturbative correlator, up to two
loops:
\begin{equation}
\wt\Pi_0^{\rm pt}(\om) = -\frac{4\Nc\om^2}
{1-2\ep}\left[\Gamma_1 +\Cf g_0^2
\frac{1-\ep}\ep
\left(\frac1{1-2\ep}\Gamma_1^2
- \frac{1-\ep/2}{1-4\ep}\Gamma_2
\right)+O\left(g_0^4\right)\right].
\label{BareCorr}
\end{equation}

For the sum rule, we need the spectral density, which is easily found as
the discontinuity of~(\ref{BareCorr}) divided by $2\pi i$. For the sake of
reliability, we have also derived it by direct use of the Cutkosky rules in
$D$ dimensions, which yields
\begin{eqnarray}
\wt\rho_0^{\rm pt}(\om) &=& \wt\rho_0^{(1)}(\om)
\left(1+\delta_2+\delta_2^*+\delta_3+O\left(g_0^4\right)\right),
\label{BareRho}\\
\wt\rho_0^{(1)}(\om) &=& \frac{2\Nc(2\om)^{D-2}}{(4\pi)^{D/2}}
\frac{\Gamma(1-\ep)}{(1-2\ep)\Gamma(1-2\ep)},
\nonumber\\
\delta_2 &=& - \frac{\Cf g_0^2(-2\om)^{-2\ep}}{(4\pi)^{D/2}}
\frac{(1-\ep)\Gamma(1+2\ep)
\Gamma(1-\ep)}
{\ep^2(1-2\ep)},
\nonumber\\
\delta_3 &=& \frac{2\Cf g_0^2(2\om)^{-2\ep}}{(4\pi)^{D/2}}
\frac{(1-\ep)(1-\ep/2)\Gamma(1-2\ep)\Gamma(1-\ep)}
{\ep^2(1-4\ep)\Gamma(1-4\ep)},
\nonumber
\end{eqnarray}
where $\wt\rho_0^{(1)}$ is the leading-order (one-loop) result, $\delta_2$
is the (complex) one-loop vertex correction~\cite{ef1} to the diagram with
the two-particle cut, and $\delta_3$ is the contribution of three-particle
cuts. Expressing $g_0^2$ in terms of $\as(\mu)$ and using $\wt Z_J$
\cite{ef1}, we obtain, by each method, the renormalized spectral function
\begin{equation}
\wt\rho^{\rm pt}(\om) = \Nc\frac{\om^2}{2\pi^2} \left[ 1 +
3\Cf\frac{\as(\mu)}{4\pi} \left( -2\log\frac{2\om}\mu
+ \frac{4\pi^2}9 + \frac{17}3  \right)+O\left(\as^2(\mu)\right)\right],
\label{RenRho}
\end{equation}
which we now compare with the results of more laborious QCD calculations.

The perturbative QCD (pseudo)scalar correlator, for any pair of quark
masses, was calculated up to two loops by one of us in~\cite{djb}. The
(pseudo)vector correlator was found in~\cite{gen}, where the results of
calculations of absorptive parts~\cite{abs} were compared and corrected.
(See also~\cite{bag}.) The pseudoscalar and pseudovector results were
obtained in the scheme with $\gamma_5$ anticommuting with all $\gamma_\mu$,
by changing the sign of a quark mass in scalar and vector results.  We may
therefore obtain EFT results from~\cite{gen,djb} by: transforming the
heavy-quark $\overline{\rm MS}$ mass to the {\em pole\/} mass $M$;
using~(\ref{EFT/QCD}) in the anticommuting $\gamma_5$ scheme; taking the
limit $M\to\infty$, with $\om$ and $m$ fixed, where $q^2=(M+\om)^2$ and $m$
is the light-quark $\overline{\rm MS}$ mass. For the spectral function this
gives
\begin{eqnarray}                                              
\wt{\rho}^{\rm pt}(\om)&=&\frac{\Nc\lambda(\om\pm m)}{2\pi^2}
\left[1+\frac{\Cf\as(\mu)}{4\pi}\left\{
\frac{6\om^2\pm6\om m-18m^2}{\lambda^2}\log\frac{\mu}{m}
\right.\right.
\nonumber\\
&+&\left.\left.
\frac{8\om}{\lambda}\left({\rm Li}\left(\frac{\om-\lambda}{\om+\lambda}\right)
-{\rm Li}\left(\frac{2\lambda}{\om+\lambda}\right)+2{\rm Li}(1)\right)
+\frac{17\om^2\pm7\om m-28m^2}{\lambda^2}
\right.\right.
\nonumber\\
&+&\left.\left.
\frac{10\om^2\pm4\om m-3m^2}{\lambda(\om\pm m)}
\log\frac{\om+\lambda}{m}
-16\log\frac{2\lambda}{m}
\right\}+O\left(\as^2(\mu)\right)\right]
\label{dlg}\\
&=&\frac{\Nc\om^2}{2\pi^2}\left[1+\frac{3\Cf\as(\mu)}{4\pi}\left\{
-2\log\frac{2\om}{\mu}+\frac{4\pi^2}{9}+\frac{17}{3}\right\}
+O\left(\as^2(\mu)\right)\right]
\nonumber\\
&\pm&\frac{\Nc\om m}{2\pi^2}\left[1+\frac{3\Cf\as(\mu)}{4\pi}\left\{
-4\log\frac{2\om}{\mu}+\frac{4\pi^2}{9}+8\right\}+O\left(\as^2(\mu)\right)
\right]
\nonumber\\
&-&\frac{\Nc m^2}{4\pi^2}\left[1+\frac{3\Cf\as(\mu)}{4\pi}\left\{
-6\log\frac{2\om}{\mu}+2\right\}+O\left(\as^2(\mu)\right)\right]
+O\left(m^3\right),
\nonumber
\end{eqnarray}
where ${\rm Li}(x)\equiv-\int_0^x dy\log(1-y)/y$ and
$\lambda\equiv(\om^2-m^2)^{1/2}$.  The agreement of~(\ref{RenRho})
with~(\ref{dlg}), as $m\to0$, confirms the matching
conditions~(\ref{EFT/QCD}) for (pseudo)scalar and (pseudo)vector currents.
The expansion of the full correlator beyond order $m^2$ involves $\log m$
terms, which must be absorbed into quark condensates~\cite{gen}.

\newpage                                       

\section{Two-loop $\qq$ contribution}
\label{QCond}

One- and two-loop diagrams for the $\qq$ contribution in any dimension $D$
and in an arbitrary covariant gauge can be systematically evaluated by the
method developed in~\cite{ef1}. We have calculated the 11 contributing
one-particle-irreducible diagrams using our REDUCE package. Combining these
with the propagator and vertex results of~\cite{ef1}, we find a
gauge-invariant $\qq$ contribution to the bare correlator, of the form
\begin{eqnarray}
\wt{\Pi}^3_0(\om)&=&\pm\frac{\lv\bar q_0q_0\rv}{2\om}\Bigl[1+P_1\Gamma_1\Cf
g_0^2+\Bigl\{(P_{{\rm A}1}\Ca+P_{{\rm F}1}\Cf)\Gamma_1^2
\nonumber\\
&+&(P_{{\rm A}2}\Ca+P_{{\rm F}2}\Cf+P_{{\rm L}2}\Tf\Nl)
\Gamma_2\Bigr\}\Cf g_0^4+O\left(g_0^6\right)\Bigr],
\label{bqq}
\end{eqnarray}
with the following coefficients, for any gauge group:
\begin{eqnarray*}
P_{1} &=&{(D-1)(D-4)\over D-3}\\
P_{{\rm A}1}&=&{(D^3-8D^2+19D-10)(D-5)\over2(D-3)^2(D-4)}\\
P_{{\rm F}1}&=&-{2(2D-9)(D-2)\over(D-3)(D-4)}\\
P_{{\rm A}2}&=&{2D^7-51D^6+552D^5-3245D^4+11064D^3-
21626D^2+22080D-8808\over4(2D-7)(D-3)^2(D-4)(D-6)}\\
P_{{\rm F}2}&=&-{D^6-20D^5+167D^4-728D^3+1704D^2-1968D+816
\over2(D-3)^2(D-4)(D-6)}\\
P_{{\rm L}2}&=&-{4(D-2)(D-4)\over(2D-7)(D-3)(D-6)}.
\end{eqnarray*}

The infinities in~(\ref{bqq}) relate the anomalous dimensions of $J$ and
$\qq$, confirming our result~\cite{ef1} for $\wt\gamma_J$. Using $\wt Z_J$,
$Z_{\bar q q}$, and $Z_\alpha$, we obtain the finite renormalized result
\begin{eqnarray}
\wt\Pi^3(\om) &=&\pm\frac{\lv\bar q q\rv_\mu}{2\om}
\Biggl\{1+\frac32\Cf\frac{\as(\mu)}\pi
\nonumber\\
&+&\Cf\left(\frac{\as(\mu)}\pi\right)^2
\Biggl[\left(\left(-\frac74+\frac{\zeta(2)}2\right)\Ca
+\left(\frac12-2\zeta(2)\right)\Cf+\Tf\Nl\right)\log\frac{2\om}\mu
\nonumber\\
&+&\left(\frac{149}{48}-\frac{3\zeta(2)}8+\frac{\zeta(3)}2\right)\Ca
+\left(\frac{11}8+\frac{5\zeta(2)}2-2\zeta(3)\right)\Cf
-\frac43\Tf\Nl\Biggr]
\nonumber\\
&+&O\left(\as^3(\mu)\right)\Biggr\}.
\label{RenQ}
\end{eqnarray}

In QCD, with $\Nl=4$, the ratio of next-next-to-leading to next-to-leading
terms is large: $(-3.00\log(2\om/\mu)+7.14) \as(\mu)/\pi$. This vanishes
for $\mu=0.185\om$, which is too low a renormalization scale to use in EFT
sum rules. However, the two-loop correction is not large by itself: about
15\% for $\mu\sim2\om\sim1$~GeV. We note that two-loop corrections to
coefficient functions of quark condensates in light-quark QCD sum rules
were calculated in~\cite{nnl}, where the corresponding ratios were found to
be even larger.


\section{Higher-dimensional condensates}
\label{Higher}

We have calculated the contributions of all quark and gluon condensates
with $d\le7$, to leading order. They are most easily found~\cite{fpg} in the
fixed-point gauge, $x_{\mu}A_{\mu}(x)=0$, in which the quark and gluon
fields have gauge-covariant Taylor expansions and external
gluons do not interact with the static quark. Using a systematic
method~\cite{gap} to reduce vacuum expectation values to a minimal basis,
we find the quark-condensate contributions
\begin{eqnarray}
\wt\Pi^q(\om) &=& \pm \frac{\lv\bar q q\rv}{2\om}
\mp \frac{\lv\bar q G_{\mu\nu}i\sigma_{\mu\nu} q\rv}{16\om^3}
- \frac{\lv\bar q J_\mu \gamma_\mu q\rv}{96\om^4}
\nonumber\\
&\pm& \frac1{96\om^5}\Biggl[6\lv\bar q G_{\mu\nu}G_{\mu\nu}q\rv
- 3\lv\bar q G_{\mu\nu}\wt G_{\mu\nu}i\gamma_5q\rv
- 6\lv\bar q G_{\mu\lambda}G_{\lambda\nu}\sigma_{\mu\nu}q\rv
\nonumber\\
&+&2\lv\bar qD_\mu J_\nu i\sigma_{\mu\nu}q\rv \Biggr]
+O\left(\frac1{\om^6}\right),
\label{HigherQ}
\end{eqnarray}
where $G_{\mu\nu}=g G^a_{\mu\nu}t^a$, $\wt G_{\mu\nu}$ is its dual,
$J_\mu=gJ^a_\mu t^a$, and $J^a_\mu=D_\nu G^a_{\mu\nu}=g\sum_q \bar
q\gamma_\mu t^a q$. The first three terms agree with~\cite{shu} and may
also be confirmed by taking the $M\to\infty$ limit of the relativistic
results in~\cite{aae,gen}.

In the fixed-point gauge, no $G^2$ term from the light quark-propagator
$S(0,x)$ survives vacuum averaging. Hence there is no one-loop
$d=4$ gluon-condensate contribution in the non-relativistic limit. We
obtain the $G^3$ contribution from the $M\to\infty$ limit of the results
of~\cite{gen} and neglect the $J^2$ term, which is commensurate with the
unknown radiative correction to the $d=6$ quark-condensate contribution
in~(\ref{HigherQ}). The gluon-condensate contribution is thus:
\begin{equation}
\wt\Pi^g(\om)=-                                            
\frac{\lv g^3f^{abc}G^a_{\lambda\mu}G^b_{\mu\nu}G^c_{\nu\lambda}
\rv}{4608\pi^2\om^4}+O\left(\frac1{\om^6}\right).
\label{HigherG}
\end{equation}

\section{Sum rule and conclusions}
\label{Concl}

To obtain a sum rule, we specialize to the SU(3) gauge group of QCD, with
$\Ca=\Nc=3$, $\Cf=4/3$, $\Tf=1/2$, and $\Nl=3$ or $4$ light flavours. From
the two-loop anomalous dimension~\cite{ef1} of $\wt J$ and the two-loop
$\beta$-function, we have
\begin{eqnarray}
\wt f_{\rm M}(\mu) = \wh f_{\rm M} \as^{-2/b}(\mu)\left(1-K\frac{\as(\mu)}\pi
+O\left(\as^2(\mu)\right)\right),
\label{RenInv}\\
K = \frac5{12} -
\frac{285-7\pi^2}{27b} + \frac{107}{2b^2},\quad b = 11-\frac23\Nl,
\nonumber
\end{eqnarray}
where $\wh f_{\rm M}$ is a renormalization-group invariant. The radiative
correction here is small: $K=0.189$ or $0.227$ for $\Nl=3$ or $4$. The
conventional QCD decay constants $f_{\rm M}$ are related
by~(\ref{EFT/QCD}) to $\wt f_{\rm M}$, as follows:
\begin{equation}
f_{\rm M} = \sqrt{\frac2M}\wt f_{\rm M}(M)
\left(1-C\frac{\as(M)}\pi +O\left(\as^2(M)\right)\right)
+O\left(\frac1M\right),
\label{fM}
\end{equation}
where $C=4/3$ for vector mesons and $C=2/3$ for pseudoscalar mesons (if a
fully anticommuting $\gamma_5$ is used). In analogy with~(\ref{RenInv}), we
also have
\begin{eqnarray*}
\lv\bar q q\rv_\mu &=& \lv\widehat{\bar q q}\rv \as^{-4/b}(\mu)
\left(1-K'\frac{\as(\mu)}\pi+O\left(\as^2(\mu)\right)\right),\\
K' &=& \frac56
-\frac{34}{3b}
+\frac{107}{b^2},
\end{eqnarray*}
where $\lv\widehat{\bar q q}\rv$ is a (negative) renormalization-group invariant.

We adopt the standard model of the continuum spectral density, setting it
equal to the perturbative one, starting at an effective threshold $\oc$.
We use the vacuum factorization approximation~\cite{svz} for $d=6$ and
$d=7$ quark condensates. For the latter, this is unreliable but serves as a
rough guide, to ensure that the contribution is kept small. The $d=5$ quark
condensate is denoted by $m_0^2\qq$. We omit the two-loop term
in~(\ref{RenQ}); to use it consistently we would need the unknown
three-loop term in $\wt\gamma_J$. Applying a Borel transform~\cite{svz,shu}
to the correlator, we obtain
\begin{eqnarray}
\wh f_{\rm M}^2 e^{-\om_{\rm M}/E}&\approx&
\frac{3E^3}{\pi^2}\left[1-\left(1+\frac{\oc}E+\frac12\frac{\oc^2}{E^2}
\right)e^{-\oc/E}\right]
\nonumber\\
&\times&\as^{4/b}(\mu)\left[1+\frac{\as(\mu)}\pi
\left(-2\log\frac{2\oc}\mu+2L\left(\frac{\oc}E\right)+2K
+\frac{4\pi^2}9+\frac{17}3
\right)\right]
\nonumber\\
&\mp&\frac{\lv\widehat{\bar q q}\rv}2\left[1+\frac{\as(\mu)}\pi
\left(2-\frac{264-14\pi^2}{27b}\right)
-\frac{m_0^2}{16E^2}+\frac{\pi\lv\as
G^a_{\mu\nu}G^a_{\mu\nu}\rv}{288E^4}
\right]\nonumber\\
&+&
\as^{4/b}(\mu)\frac{\pi\as\qq^2-
\frac3{256\pi^2}
\lv g^3f^{abc}G^a_{\lambda\mu}G^b_{\mu\nu}
G^c_{\nu\lambda}\rv}{324E^3},
\label{SumRule}\\
L(x)&=&\frac{\log x+\gamma+E_1(x)
+\frac12(3+x)e^{-x}-\frac32}
{1-\left(1+x+\frac12x^2\right)e^{-x}}
\nonumber\\
&=&\left\{
\begin{array}{ll}
\frac13+\frac1{16}x+\frac{43}{1600}x^2+O\left(x^3\right),&\,x\ll1,\\[3pt]
\log x+\gamma-\frac32+O\left(e^{-x}x^2\log x\right),&\,x\gg1,
\end{array}
\right.
\nonumber
\end{eqnarray}
where the upper (lower) sign is for ground (excited) state mesons, $\gamma$
is Euler's constant, and $E_1(x)=\int_x^\infty dy\;e^{-y}/y$.

The left-hand side of sum rule~(\ref{SumRule}) is a renormalization-group
invariant; so is the right-hand side, to the given order in $\as(\mu)$.
Numerical investigation, over a suitable range of the Borel variable $E$,
reveals that all terms may be kept under control, save one. The glaring
exception is the perturbative correction, which invariably exceeds
$(-2\log(2\oc/\mu)+11.1)\as(\mu)/\pi$. For $\mu\sim\oc\sim1$~GeV, this
is of order 100\%, an order of magnitude greater than in the QCD sum rule
for $\rho$~\cite{svz} and, in our opinion, far too large for one to have
any confidence in the neglect of unknown three-loop terms. An alternative
way to express this problem is to say that fastest apparent convergence,
with a vanishing next-to-leading correction, occurs at a scale
$\mu<0.008\oc\sim10$~MeV, an order of magnitude smaller than
$\Lambda_{\overline{\rm MS}}$. The radiative correction to the
quark-condensate contribution, on the other hand, is acceptably small:
around 15\%.

We conclude that non-relativistic sum rules for mesons with a heavy quark
are intrinsically unreliable. Whilst EFT solves the hybrid log problem, it
also leaves one with a unique perturbative correction whose numerical size
precludes any reliable determination of $\wh f_{\rm M}$ and hence any
estimate of $f_{\rm B}$.             

\setlength{\parindent}{0cm}
{\bf Acknowledgements\ } AGG thanks V.~S.~Fadin for advice on dispersive
calculations. DJB thanks B.~L.~Ioffe, C.~A.~Dominguez, N.~Paver and
H.~G.~Dosch for helpful discussions. A grant from SERC to enable our
collaboration is gratefully acknowledged.

\newpage                             

\raggedright

\end{document}